**Urban and non-urban contributions to the social cost of carbon**

Francisco Estrada[1,2,3*], Veronica Lupi[2,4], Wouter Botzen[2], Richard S.J. Tol[5,2,6,7,8,9,10]

[1]*Instituto de Ciencias de la Atmósfera, Universidad Nacional Autónoma de México, Mexico City 04510, México;* [2]*Institute for Environmental Studies (IVM), Vrije Universiteit, Amsterdam 1081 HV, the Netherlands;* [3]*Programa de Investigación en Cambio Climático, Universidad Nacional Autónoma de México, Mexico City, Mexico;* [4]*Fondazione Eni Enrico Mattei, Milan;* [5]*Department of Economics, University of Sussex, UK;* [6]*Department of Spatial Economics, Vrije Universiteit, Amsterdam, the Netherlands;* [7]*Tinbergen Institute, Amsterdam, the Netherlands;* [8]*CESifo, Munich, Germany;* [9]*Payne Institute of Public Policy, Colorado School of Mines, Golden, CO, USA;* [10]*College of Business, Abu Dhabi University, UAE.*

*\*Corresponding author*

Email: *feporrua@atmosfera.unam.mx*

**The social cost of carbon (SCC) serves as a concise gauge of climate change's economic impact, often reported at the global and country level. SCC values are disproportionately high for less-developed, populous countries. Assessing the contributions of urban and non-urban areas to the SCC can provide additional insights for climate policy. Cities are essential for defining global emissions, influencing warming levels and associated damages. High exposure and concurrent socioenvironmental problems exacerbate climate change risks in cities. Using a spatially explicit integrated assessment model, the SCC is estimated at USD\$137-USD\$579/t$CO_2$, rising to USD\$262-USD\$1,075/t$CO_2$ when including urban heat island (UHI) warming. Urban SCC dominates, with both urban exposure and the UHI contributing significantly. A permanent 1% reduction of the UHI in urban areas yields net present benefits of USD\$484-USD\$1,562 per urban dweller. Global cities have significant leverage and incentives for a swift transition to a low-carbon economy, and for reducing local warming.**

Limiting global warming below 2°C relies upon a carbon budget of 580 Gt$CO_2$, and implies reaching carbon neutrality in approximately thirty years[1]. Nationally Determined Contributions (NDCs) are key components of the Paris Agreement[2], with nations' progress continuously monitored[3]. The latest estimates suggest strict compliance of current NDCs will mitigate climate change risks, but an 80% increase in the average emission reduction is needed to avoid exceeding 1.5°C globally[1,4,5]. However, the decarbonization of the economy has proven highly complex, and policy interventions are slowly implemented. Current per-capita average global emissions exceed by four tonnes of carbon dioxide equivalent the yearly budget for the 1.5°C target[6].

Climate and weather hazards intensified in the past years, affecting economies and societies through extreme events such as heatwaves, floods, droughts, and storms, as well as chronic impacts like decreasing crop yields, diseases, and rising sea levels. Beside imposing huge costs on populations, these extreme events worsened inequalities and conflicts[7–9], as exposure, hazard and vulnerability are heterogeneous among regions[10–12]. In response, the 27th Conference of the Parties advocated the establishment of a fund for *loss and damage*[13]. This initiative is grounded in the "polluters pay" principle, and aims at establishing compensation schemes for vulnerable countries with small contributions to current global warming[14].

Several metrics have been proposed to gauge the economic damages that climate change would cause under different greenhouse gases (GHG) emissions scenarios. The SCC is a key metric that focuses on marginal damages. The SCC estimates the economic costs an additional tonne of $CO_2$ released into the atmosphere would produce in terms of impacts on natural and human systems. Despite several shortcomings of SCC to realistically reflect the true cost of carbon[15–17], and of the models that produce them[8,18,19], it remains a convenient summary metric for advising policy making[20]. The SCC is a benchmark used in cost-benefit analysis to estimate the monetary benefit of emissions' cuts by mitigation policy proposals[21–23] and it guides governments in optimally pricing carbon emissions. Recently, the US administration re-established the Interagency Working Group (IWG) on Social Cost of GHGs emissions[22], to inform the US climate policies. A recent study[24] that uses a comprehensive modelling approach suggests that the SCC would be between USD\$44/t$CO_2$ USD\$413/t$CO_2$, with a central estimate of USD\$185/t$CO_2$. Estimates in the literature can have a much wider range for the SCC including from \$10/t$CO_2$ to \$1,000/t$CO_2$ and even from -\$200/t$CO_2$ to \$30,000,000/t$CO_2$, depending on the assumptions and discount rate applied[17,25,26].

The SCC is also used to evaluate the regional and national distribution of climate impacts and inequality concerns. These studies agree on large SCC heterogeneity among regions and countries such that those with large shares of global GDP and/or high sensitivity to climate change bear the largest shares[23,25]. These countries typically include India, China, the EU and the US. A recent study[27] develops national-level damage functions that account for differences in income level and annual temperature. It finds that the national SCC is higher in less-advanced countries with large populations than in the richest ones. Consequently, poor countries would have the highest incentives to set the highest carbon prices if acting in their self-interest. In addition, estimates of the *local* cost of carbon are being proposed, defined as the social cost of carbon if everyone in the world were like the people in a particular locale[28].

Urban agglomerations and cities are key in determining the global emissions pathway. Cities consume about 78% of the world's energy and are responsible for almost 8 of every 10 tonnes of GHG emitted to the atmosphere. The top 100 emitting urban areas accounting for 18% of the global carbon footprint[29–31]. About 56%-85% of global population lives in cities and produce close to 80% of global GDP[30–33]. These trends are projected to accelerate in the coming decades[31]. The UHI effect occurs when natural landscapes are replaced by denser, higher thermal capacity materials which lead to a local energy imbalance and changes in local climate[34–36]. The UHI effect produces negative impacts such as intensified heat waves[37], higher energy consumption[38], lower labor productivity[39], increases in human health risks and discomfort[37,40], higher water demand[41] and increased air and water pollution[42,43]. It has been estimated that in large cities the UHI effect can add up to 4ºC of warming and that when this phenomenon is included in the assessment of the economic impacts of climate change, global losses increase by at least twofold[4,44,45]. At the same time, cities have a disproportionate influence on national and global climate policy and can implement relatively costless local adaptation measures to reduce UHI warming[44,46,47].

Previous global to national estimates of the SCC have omitted the UHI effect and have not investigate the contributions from urban areas to these figures. Here these limitations are addressed, and global and regional urban and non-urban SCC estimates are provided. Moreover, urban SCC values are separated into their exposure and UHI effects. For this purpose, the CLIMRISK integrated assessment model[4] is used (see Methods). This is a spatially explicit integrated assessment model that allows modelling the UHI effects and estimating the economic impacts of both global and local (UHI) warming for urban areas. CLIMRISK uses spatially explicit climate and socioeconomic data and several damage function specifications, to reflect the uncertainty surrounding SCC estimates. CLIMRISK aggregates results in 13 world regions defined in Table S15. Some of the damage functions (DF) in this model are path-, scenario-, and time-dependent, and vary both at the regional and grid-cell level. In the results presented here four types of regional/local DF are used (see Methods): the *R* regional DF which are based on the RICE2010 model[48], but that is driven by local temperature projections and that has been updated to match the calibration of the latest DICE DF[49], as well as that of Weitzman[50] which aims to represent catastrophic damages from climate change; the *RU* regional and local DF that include the economic impacts from local climate change, particularly the urban heat island effect; *RP* regional DF that include the persistence of impacts from climate change in the economy, and; *RPU* regional and local DF that include both the economic impacts of urban warming and the effects of the persistence of impacts from climate change on the economy. Results in the main text are presented for the SSP585 scenario, and the Supplementary Information offers results for the SSP370, SSP245 and SSP126 scenarios.

**Results**

*Global and regional SCC estimates for different damage functions*

The global SCC obtained using the *R* functions amounts to $137/tCO2 (Table 1). Africa, India, China and OASIA account for about 63% of this value, while the EU, US and OHI only for 22%. However, the omission of local warming in urban agglomerates and the dynamics of impacts has been shown to produce large downward biases in the assessment of the economic costs of climate change[4,44,51,52]. The global SCC value rises to $262/tCO$_2$ with the effects of the UHI in urban areas (RU). When including persistence of impacts (*RP*) and both persistence and UHI (*RPU*), the SCC values are $579/tCO$_2$ and $1,074/tCO$_2$, respectively. In all cases the distribution of SCC among the different regions is highly heterogeneous. The regions with the highest SCC values are Africa ($28/tCO$_2$ (*R*), $293/tCO$_2$ (*RPU*)), India ($25/tCO$_2$ (*R*), $190/tCO$_2$ (*RPU*)), and China ($17/tCO$_2$ (*R*), USD$190/tCO$_2$ (*RPU*)). Figure S1 shows the SCC estimated for different SSP scenarios and illustrates the large influence GDP and population assumptions have on this metric (Tables S1-S4). It is notable that under high climate scenario and low economic growth (SSP370), the global SCC values are lower than those obtained in a considerably lower warming/higher economic growth scenario (SSP245). In contrast, if instead of the SSP3 economic scenario the SSP5 is used, the SSP570 gives very similar results to those of the SSP585.

**Table 1: Social cost of carbon for different damage functions in CLIMRISK.**

| Region | R | RP | RU | RPU |
|---|---|---|---|---|
| US | $12.24 (8.94%) | $28.88 (4.99%) | $25.54 (9.73%) | $60.31 (5.61%) |
| EU | $14.35 (10.48%) | $27.61 (4.77%) | $28.78 (10.97%) | $55.43 (5.16%) |
| Japan | $2.57 (1.88%) | $6.10 (1.05%) | $5.18 (1.97%) | $12.31 (1.15%) |
| Russia | $1.72 (1.26%) | $7.59 (1.31%) | $4.9 (1.87%) | $21.63 (2.01%) |
| Eurasia | $1.63 (1.19%) | $5.04 (0.87%) | $3.65 (1.39%) | $11.28 (1.05%) |
| China | $17.48 (12.78%) | $78.02 (13.47%) | $44.65 (17.02%) | $199.52 (18.56%) |
| India | $24.81 (18.13%) | $107.39 (18.54%) | $44.03 (16.78%) | $190.44 (17.72%) |
| MEAST | $6.29 (4.6%) | $27.35 (4.72%) | $12.52 (4.77%) | $54.43 (5.06%) |
| Africa | $27.97 (20.44%) | $192.54 (33.24%) | $42.43 (16.17%) | $292.63 (27.22%) |
| LAM | $6.18 (4.52%) | $16.56 (2.86%) | $11.78 (4.49%) | $31.57 (2.94%) |
| OHI | $4.17 (3.05%) | $9.84 (1.7%) | $8.55 (3.26%) | $20.19 (1.88%) |
| OASIA | $15.74 (11.5%) | $67.81 (11.71%) | $26.96 (10.27%) | $116.00 (10.79%) |
| MX | $1.67 (1.22%) | $4.47 (0.77%) | $3.43 (1.31%) | $9.19 (0.86%) |
| WORLD | $136.83 (100%) | $579.19 (100%) | $262.4 (100%) | $1074.94 (100%) |

Numbers in parenthesis indicate the fraction of global SCC for each region. Figures are expressed in US2005 dollars. 1.5% consumption discount rate. DICE2016 global DF. Global aggregates are not equity-weighted.

The regional share of the global SCC is highly influenced by the assumptions made about urban effects and persistence of impacts. While for all regions the estimated SCC values increase as the UHI, persistence and both factors are included, the fraction of the global SCC that each region accounts for are correlated to their development level (see Tables S5-S8 for the Weitzman global DF and Tables S9-S12 for a 3% discount rate). For regions such as the EU, US, Japan, and OHI including persistence lowers in about 44%-55% their contribution to the global SCC, suggesting that, in comparison, other regions have more limited resilience and capacities to deal with climate shocks. On the contrary, for all regions, the fraction of global SCC increases when urbanization (*RU*) is included, except Africa (-4.03 percentage points(pp)), India (-1.40pp) and OASIA (-1.20pp). That is, the increases in costs due to urban effects and exposure in these developing regions tend to be smaller than in those that are highly developed. The increase in the fraction of global damages due to the inclusion of urban warming is largest in China (4.24pp), followed by the US (0.79pp), Russia (0.61pp), and EU (0.48pp). The fraction of SCC increases steeply for Africa (6.78pp) and China (5.78pp) when including persistence and urban effects.

*Urban and rural contributions to the SCC at the regional and global levels*

Given the importance of the UHI effects for the global and regional SCC values, we separate the urban (*u*) and non-urban (*nu*) contributions to these estimates (see Methods). Moreover, urban damages are further decomposed into exposure and urban-specific effects. As shown in Table 2, the SCC values for urban areas are much higher than those for non-urban ones. At the global level, the estimated urban SCC value is between 7 and 14 times larger than that of non-urban areas depending on the comprehensiveness of the damage function.

The contribution of urban areas to the total global SCC is about 93%, with a regional range varying from 85% (Eurasia) to 99.7% (India). On average, the regional urban SCC values are 41.4 times as large as those for non-urban, with some regions such as India and China reaching 357 and 47 times, respectively. In particular, the global urban SCC estimates are $245/tCO$_2$ (*RU*) and $1,006/tCO$_2$ (*RPU*).

**Table 2: Non-urban and urban SCC and the contributions of exposure and urban characteristics to urban SSC.**

| Region | R,RU damage functions | | | | | RP,RPU damage functions | | | | |
|---|---|---|---|---|---|---|---|---|---|---|
| | Non-urban | Urban | Urban-noUHI | Exposure | UHI+int | Non-urban | Urban | Urban-noUHI | Exposure | UHI+int |
| US | $3.14 | $22.40 | $9.09 | $5.95 | $13.30 | $7.41 | $52.90 | $21.48 | $14.07 | $31.42 |
| EU | $2.22 | $26.57 | $12.13 | $9.92 | $14.44 | $4.26 | $51.17 | $23.35 | $19.09 | $27.82 |
| Japan | $0.15 | $5.03 | $2.42 | $2.27 | $2.61 | $0.36 | $11.95 | $5.74 | $5.38 | $6.22 |
| Russia | $0.54 | $4.36 | $1.18 | $0.64 | $3.18 | $2.38 | $19.24 | $5.21 | $2.83 | $14.03 |
| Eurasia | $0.56 | $3.09 | $1.08 | $0.52 | $2.02 | $1.72 | $9.56 | $3.32 | $1.60 | $6.24 |
| China | $0.93 | $43.72 | $16.55 | $15.62 | $27.17 | $4.17 | $195.34 | $73.85 | $69.68 | $121.49 |
| India | $0.12 | $43.90 | $24.69 | $24.57 | $19.21 | $0.53 | $189.91 | $106.86 | $106.33 | $83.05 |
| MEAST | $0.93 | $11.59 | $5.36 | $4.44 | $6.22 | $4.02 | $50.42 | $23.33 | $19.32 | $27.09 |
| Africa | $4.98 | $37.71 | $22.99 | $18.00 | $14.72 | $34.31 | $260.06 | $158.23 | $123.92 | $101.83 |
| LAM | $1.37 | $10.41 | $4.81 | $3.44 | $5.60 | $3.68 | $27.91 | $12.88 | $9.20 | $15.03 |
| OHI | $0.92 | $7.64 | $3.25 | $2.33 | $4.39 | $2.16 | $18.05 | $7.68 | $5.51 | $10.38 |
| OASIA | $1.17 | $25.80 | $14.57 | $13.39 | $11.24 | $5.05 | $111.04 | $62.76 | $57.71 | $48.28 |
| MX | $0.30 | $3.13 | $1.37 | $1.07 | $1.76 | $0.81 | $8.38 | $3.66 | $2.85 | $4.73 |
| WORLD | $17.34 | $245.35 | $119.49 | $102.14 | $125.86 | $70.86 | $1,005.95 | $508.34 | $437.48 | $497.61 |

1.5% discount rate. Figures are expressed in US2005 dollars. *Non-urban* represents the SCC from non-urban areas; *Urban* shows the SCC value in urban areas when the UHI effect is accounted for; *Urban-exposure* provides the SCC value in urban areas but when the UHI effect is not considered; *Inc-exposure* separates the impact on SCC from increased exposure in cities; *UHI-int* shows the contribution to the SCC from including the UHI effect plus its interactions with global climate change (see Methods).

The global non-urban SCC is $17/tCO$_2$, and $71/tCO$_2$ for *R* and *RP*, respectively. The regions with the largest non-urban SCC are Africa (*R*: $5/tCO$_2$; *RP*: $34/tCO$_2$), the US (*R*: $3/tCO$_2$; *RP*: $7/tCO$_2$), the EU (*R*: $2/tCO$_2$; *RP*: $4/tCO$_2$), LAM (*R*: $1/tCO$_2$; *RP*: $4/tCO$_2$), OASIA (*R*: $1/tCO$_2$; *RP*: $5/tCO$_2$), China (*R*: $1/tCO$_2$; *RP*: $4/tCO$_2$) and MEAST (*R*: $1/tCO$_2$; *RP*: $4/tCO$_2$). The global urban SCC estimates are much higher, and the heterogeneity in SCC values at the regional level is considerable. The six regions with largest urban SCC are India, China, Africa, EU, OASIA, and US account for more than 80% of global SCC).

*Disentangling the effects of urban exposure from those of local climate change and urban interactions*

One of the main drivers of economic losses in both the literature of extreme events and climate change is exposure. Typically, cities are subject to higher exposure than non-urban areas, due to the concentration of population and economic activities, which are expected to contribute considerably to the SCC values. We explore the composition of damages in urban areas by estimating the SCC values if no local climate change occurred, and if the sensitivity to global climate change in those areas would be the same as for the whole region (see Methods).

The global SCC in urban areas without considering the UHI effects (*Urban-exposure*) is USD$119/tCO$_2$ ($508/tCO$_2$ including persistence), about 7 times as large as the SCC for non-urban areas (*Non-urban*). That is, the differences in exposure (*Inc-exposure*) alone amount to $102/tCO$_2$ and $437/tCO$_2$ for the damage functions that do and don't include persistence of impacts, respectively. In all cases, increased exposure in urban grid cells leads to higher losses for every region. However, the estimates of the contribution of increased exposure in urban areas are highly heterogeneous among regions. Increased exposure in urban areas (*Inc-exposure*) represents 56% and 50% of the total SCC values for India and OASIA, respectively, and 39% for the global SCC value. The lowest values occur in Russia, Eurasia and the US.

Globally, the contribution of the UHI and interaction effects with global climate change (*UHI-int*) amounts to $126/tCO$_2$ *(RU)* and $498/tCO$_2$ (RPU). The *UHI-int* contribution to the SSC is similar in magnitude to that of the urban exposure. The largest contributions from *UHI-int* are in developing regions such as China ($27/tCO$_2$ *(RU)*; $121/tCO$_2$ *(RPU)*), India ($19/tCO$_2$ *(RP)*; $83/tCO$_2$ *(RPU)*), Africa ($15/tCO$_2$ *(RP)*, $102//tCO$_2$ *(RPU)*), but also in the US and the EU with values close to $14/tCO$_2$ *(RP)* and $30/tCO$_2$ *(RPU)*.

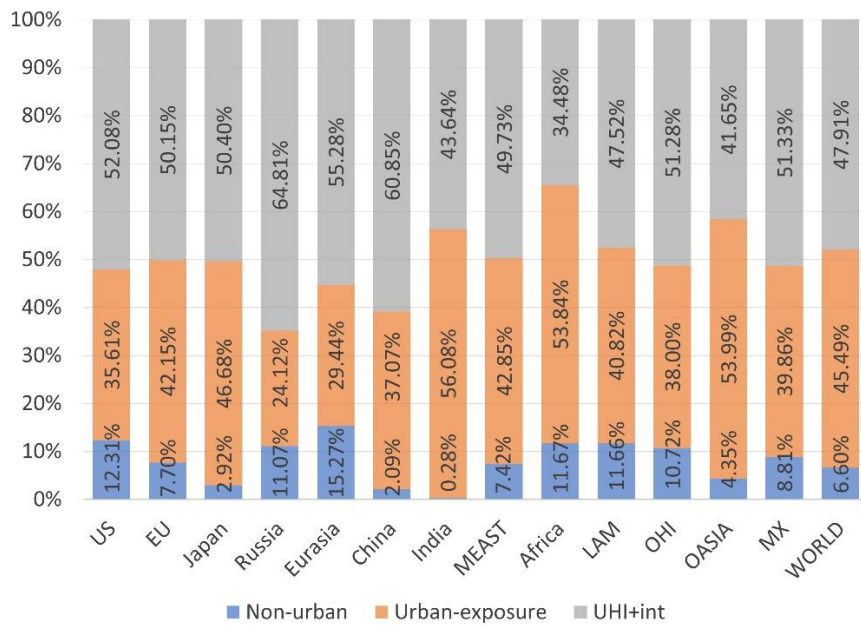

**Figure 1. Contributions of non-urban areas, urban exposure, and UHI plus interactions in urban areas to regional and world SCC.** Numbers are in percent of total SCC and add up to 100%.

In general, the contribution of non-urban areas to the total SCC is small, about 7% at the global level, and weighs less than 10% in almost half of the regions. The regions with the largest contributions from non-urban areas are Eurasia (15.27%), the US, LAM, Africa, Russia (about 12%-13%), followed by OHI (10.72%). Regarding Mexico, EU and MEAST, the contribution of non-urban damages to their SCC is larger than the global estimate, while for India, China and Japan, these numbers are very small (0.28%- 2.92%).

At the global level, *Urban-exposure* and *UHI-int* account for about 48% and 46% of the total SCC, respectively. The regions in which urban exposure contributes more than 50% to their total SCC are India, Africa and OASIA. In contrast, the UHI plus interactions effects dominate (>50%) in Russia (64.81%), China (60.85%), Eurasia (55.28%), US (52.08%), Mexico (51.33%). These effects are larger than the global estimate also for the EU, OHI, and MEAST. Africa, by far, has the lowest contribution of UHI effects to its regional SCC. These figures show that urban areas dominate the SCC, both at the global and regional levels. The reported SCC values and decomposition are the product of complex interplay between population and GDP growth and concentration, as well as warming, all at the grid-cell level. Different development paths defined by the SSP lead to different contributions to regional SCC, with regions like Africa, OHI, China and the US showing large variability (Table S13).

*3.2.2 Evaluating the marginal social benefits of mitigating the UHI*

A new metric defined as the Social Cost of the UHI (SCUHI) is applied for estimating the benefits the average urban dweller would experience from a marginal reduction in UHI. It is calculated as the variation in the present value of damages per average urban dweller from local and global climate change produced by decreasing the UHI by 1% in global urban areas over a prolonged period (e.g., 2100; see Methods). Note that keeping UHI reductions at 1% implies that the absolute mitigation of the UHI in terms of degree Celsius is likely to increase over time. This metric can be useful in climate urban policy design as it aims at providing a quantitative estimate of the expected benefits of local policies to reduce the UHI effect, such as expanding green nature, cool roofs and cool pavements.

Table 3 : Estimates of the social cost of the UHI for the RU and RPU damage functions.

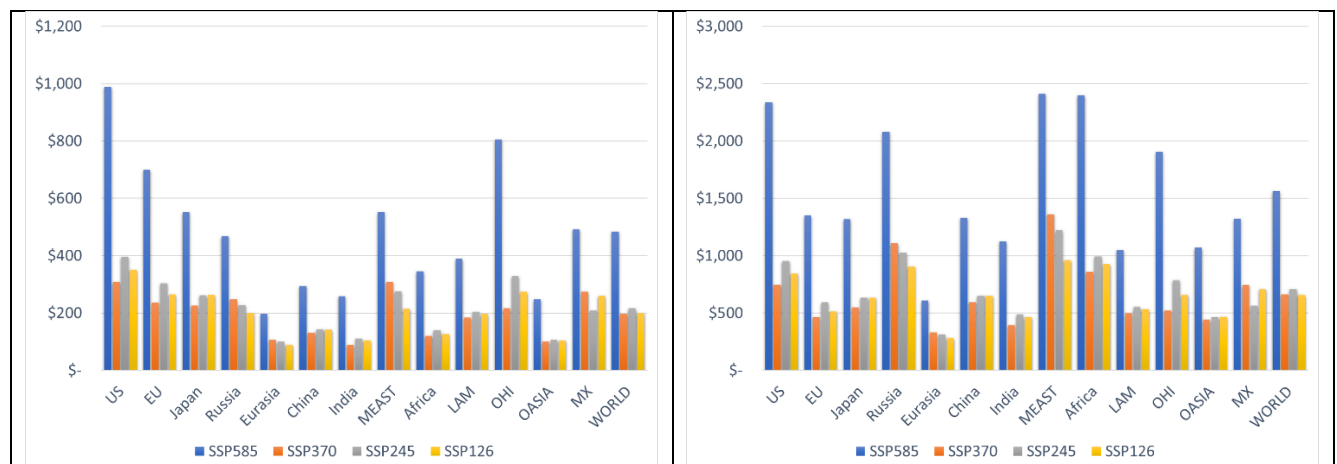

**Figure 2. Social Cost of the UHI estimates for 13 world regions and four SSP scenarios.** Panel a) shows SCUHI estimates for the RU damage function, while panel b) for the RPU damage function. Figures are in US2005 dollars. 1.5% discount rate. DICE2016 global DF.

Globally and for the SSP585, the total benefits from reducing the UHI by 1% are estimated in the range of $2,344 billion (*RU*) and $9,132 billion (*RPU*). These benefits are large and comparable to 30%-40% of the present value of the accumulated climate change losses countries such as Mexico would face this century under a high-emissions scenario (Estrada and Botzen, 2021).

The distribution of benefits is highly heterogeneous among regions and varies according to the SSP scenario and type of damage function (Figure 2). Globally, under the SSP585, the direct benefits of reducing the UHI for the average urban dweller range between $484 (*RU*) and $1,562 (*RPU*). Under the *RU*, the largest SCUHI values occur in the highest income regions, such as the US ($988), OHI ($805), the EU ($700) and Japan ($552). The lowest SCUHI values are in the least developed regions such as Eurasia ($196), OASIA ($247), India ($259), and Africa ($346). The distribution of SCUHI among regions is more mixed with the RPU damage functions: the largest SCUHI values occur in the US ($2,338), MEAST ($2,410), Africa ($2,398), Russia ($2,081), and OHI ($1,905), while the lowest SCUHI values are in EURASIA ($608), LAM ($1,051), OASIA ($1,072), and India ($1,124). For the other SSP scenarios considered, the SCUHI values are considerably lower but illustrate that higher SCUHI values can be obtained for lower warming scenarios depending on their socioeconomic assumptions and their effects on local warming and exposure (Figure 2).

**Discussion**

The consequences of unabated climate change at the global and regional scales are shown to be substantially higher than previously estimated. Omitting persistence of shocks, as is done in most current IAMs, implies that climate change damages only affect the period when they occur and have no influence on future periods. This is equivalent to assuming perfect resilience and autonomous adaptation which, in addition, is costless[51]. The warming in local temperatures in large cities is comparable to that estimated for global temperatures under a high-emissions scenario at the end of this century. By itself, the UHI effect has large economic costs, and global climate change can considerably amplify them[44]. By decomposing the SCC into urban and non-urban areas, results show that impacts in urban areas dominate the SCC, both at the global and regional levels. Depending on the region, UHI effects or persistence can dominate the increase in damages.

Cities are one of the main contributors to global greenhouse gas emissions and results show that they are also likely to experience the largest SCC. The decomposition of SCC into urban and non-urban components shows that about 93% of the global SCC comes from urban areas. Previous studies suggest that poorer regions would have the largest incentives to support the highest $CO_2$ prices, as they would be the ones that suffer most climate change impacts. The detailed treatment of urban areas in CLIMRISK allows for the decomposition of urban and non-urban SCC and shows that global and regional urban population is the most affected. Urban dwellers have the largest incentives for reducing $CO_2$ emissions and promote higher $CO_2$ prices. This is of particular importance since large cities have economic power and political capacity to transition to lower emissions development paths. Moreover, they wield considerable influence for enhancing mitigation efforts at national level and advocating for more ambitious international targets in global climate negotiations. A more stringent mitigation of greenhouse gases is in the best interest of urban regions in the world, particularly those in the most developed regions. In addition, local adaptation strategies to reduce the damages from global and local climate change have low political and economic implementation costs. The SCUHI metric indicates that the benefits for urban dwellers from these measures are substantial.

## Methods

*CLIMRISK description*

CLIMRISK is a new generation IAM of the climate and economy which allows projecting the economic impacts and risks of climate change at the global, regional and spatially explicit (0.5ºx0.5º) scales[4]. It is structured in four main modules: 1) Socioeconomic, which represents exposure in terms of population an GDP; 2) Climate, it is divided in three sections which are a global climate model, a general circulation models emulator to produce spatially explicit temperature and precipitation projections, and a third section that projects UHI intensity in urban areas as a function of population counts; 3) Economic damages, which contains sets of global, regional and grid scale-defined damage functions (DF) to map increases in temperature to losses in GDP. Some of these DFs are time-, path and scenario-dependent (*RU*, *RP*, *RPU*); 4) Risk evaluation, it combines the results from the climate and economic damages modules to produce uni- and multi-variate risk indices. 3 shows a stylized description of the core modules of CLIMRISK. The model aggregates results in 13 regions which are described in Table S15. The following paragraphs present the improvements introduced in CLIMRISK in this paper. the reader is referred to the complete and detailed description of the model contained in the Supplementary Information of Estrada and Botzen[4].

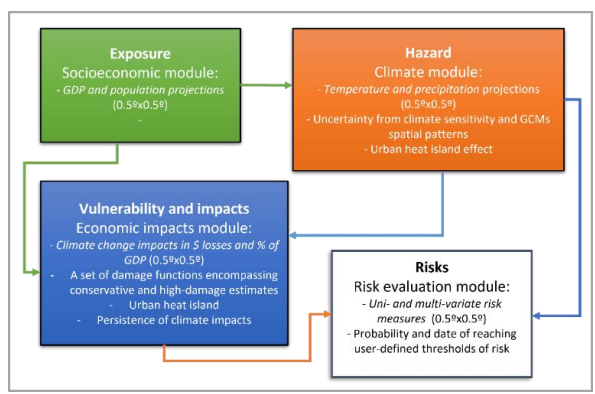

Figure 3. Schematic description of CLIMRISK's model structure. Adapted from Estrada and Botzen[4].

*Improvements in CLIMRISK's climate module*

Th climate module in CLIMRISK has been updated to reflect some of the recent advances reported in the latest assessment report (AR6) of the IPCC[53]. CLIMRISK uses a triangular probability distribution to represent the uncertainty of the climate sensitivity parameter in MAGICC6[54,55], a reduced complexity climate model. The parameters of the triangular distribution have been updated to reflect the very likely range in the AR6. The triangular distribution in the current version of CLIMRISK is specified with 2ºC and 5ºC as the lower and upper limits, respectively, with a most likely value of 3ºC.

CLIMRISK uses the pattern scaling technique[56–59] to generate spatially explicit annual temperature and precipitation scenarios (see ref[4]). Currently, the climate module in CLIMRISK contains a pattern scaling library of 37 Atmosphere-Ocean General Circulation Model (AOGCM) included in the Coupled Model Intercomparison Project (CMIP6). These patterns are produced using ordinary least squares regression as described in the original CLIMRISK paper[4] and others[58,60,61]. Table S16 provides a list of names of the AOGCMs included in the new version of CLIMRISK.

*Extension of damage functions in CLIMRISK*

In this version of the model, a variety of global DFs can be integrated into the analysis of climate impacts at the regional and local scales. In particular, in this paper CLIMRISK includes the DICE2016 global DF[23], and the global DF proposed by Weitzman (2009) which is a highly nonlinear function of global temperatures (results in Supplementary Information). These two global DF were chosen because they represent contrasting views about the seriousness of climate change impacts to the global economy. The first DF is commonly considered as conservative, and the latter focusses much more on the possibility of extreme outcomes and climate catastrophes from global warming.

Here we use CLIMRISK to downscale the global aggregated economic impacts proposed by these global DF using the regional DF from RICE, which are adjusted in CLIMRISK to produce grid-cell level estimates (*R* DF) which are consistent with the chosen global DF. In addition, features in CLIMRISK such as the inclusion of the UHI effect and persistence can be incorporated, given rise to modified versions of the *RU* and *RPU* DFs.

The downscaling procedure consists in calculating a scaling factor from the global damages $I_t^{DF^*}$ calculated with the target global DF (e.g., DICE2016) and the globally aggregated $I_t^R$ obtained using the *R* DF in CLIMRISK:

$$S_t = I_t^{DF^*} \big/ I_t^R$$

This scaling factor is applied to the grid-cell projections obtained with the different DFs included in CLIMRISK, namely *R*, *RU*, *RPU*.

*Definition of urban areas*

In the previous version of CLIMRISK[4], urban areas were declared in grid cells with population counts larger than 1 million inhabitants. This calibration was found to be very conservative as it led to underestimating the global urban population (about 40%) and GDP (about 50%). In the current version of the model, the population count threshold is calibrated to be better aligned with observed estimates of the percentages of global urban population and GDP. These estimates are uncertain, but urban population is estimated to represent between 56% and 85% of the total global population[33,62]. The population count threshold per grid cell to declare a urban area was chosen to be 250,000 inhabitants, which leads to population counts that represent about 62% of global population and about 78% of global GDP, which is aligned to estimates in the literature[30,32,63].

*The UHI effect and the urban damage functions in CLIMRISK*

Consider a Nordhaus- type damage function $D(T) = \alpha T^t$, where $D$ are the projected losses in GDP (%) which are a function of temperature change $T$. Including local temperature change caused by the UHI effect in urban areas results in:

$$D(T) = D_R(T_R) + D_U(T_U) = D_R(T_{GHG}) + D_U(T_{GHG} + T_{UHI}) = \alpha_R T_{GHG}^2 + \alpha_U(T_{GHG} + T_{UHI})^2 =$$

$$\alpha T_{GHG}^2 + 2\alpha_U T_{GHG} T_{UHI} + \alpha_U T_{UHI}^2 = D(T_{GHG}) + 2\alpha_U T_{GHG} T_{UHI} + \alpha_U T_{UHI}^2$$

Where $D_R$ and $D_U$ represent losses in rural and urban areas, $T_{GHG}$ is the change in temperature due to global warming (i.e., anthropogenic forcing), and $T_{UHI}$ is local warming in urban areas due to UHI. In words, total damage equals rural damage plus urban damage. Urban temperature change equals rural warming plus urban warming. If the damage is a power (=2) function in temperature, total damage equals total damage of greenhouse warming *plus* urban damage of urban warming *plus* the interaction of urban and greenhouse warming. Ignoring the urban heat island effect, $T_{UHI} = 0$, as is done in other integrated assessment models downward biases economic losses[4,44].

*Persistence of economic impacts of climate change*

CLIMRISK's DF include the temporal dynamics of climate change impacts. The DF in most IAMs assume that climate impacts only affect the period when they occur have no persistence. The omission of impact dynamics can be interpreted as assuming an autonomous, unlimited, costless and effective reactive adaptation capacity[51]. Persistence of socks is a stylized fact of macroeconomic time-series that has been long documented in the literature[64]. Similarly, impacts in human and natural systems tend to dissipate only after a certain period of time. Most of IAMs and assessments of the economic impacts of climate change ignore this feature which can lead to large biases in estimates[8,51,65–67]. In CLIMRISK, the persistence of impacts is incorporated into the projection of the costs of climate change for each region by means of the following equation:

$$I_{r,t}^p = Y_{r,t}D_{r,t} + \varphi I_{r,t-1}^p$$

for $t > 1$ and $I_{r,t=1}^p = Y_{r,t=1}D_{r,t=1}$

with $I_{r,t}^p$ representing the aggregated economic losses when the effects of persistence for region r are accounted for, and the persistence parameter is restricted to $0 \leq \varphi \leq 1$. The regional values of $\varphi$ are reported in Table S21 of ref[51].

*Modelling of a $CO_2$ pulse in global temperature*

The estimation of the SCC requires introducing the effects on temperatures from a pulse in CO2. In this version of CLIMRISK we adopt the approximation proposed by (Ricke and Caldeira, 2014) which is physically sound and easy to implement in IAMs. Following the authors, the effect of a pulse of 1GtC on global temperatures can be approximated using three-exponential functions:

$$\Delta T_t = -(a_1 + a_2 + a_3) + a_1 e^{-\frac{(t-t_0)}{\tau_1}} + a_2 e^{-\frac{(t-t_0)}{\tau_2}} + a_3 e^{-\frac{(t-t_0)}{\tau_3}}$$

Where $a_1 = -2.308$, $a_2 = 0.743$, $a_3 = -0.191$, $\tau_1 = 2.241$, $\tau_2 = 35.750$, $\tau_3 = 97.180$ (see[68]). The units of $\Delta T_t$ are mK/GtC. The year of the pulse ($t_0$) is chosen to be 2010.

*Decomposing the SCC into urban and non-urban contributions*

We use CLIMRISK's explicit spatial resolution and special modelling of urban areas to calculate SCC values and to decompose these figures into dominantly urban and non-urban (mixed and rural) contributions. Moreover, exploiting the set of damage functions included in CLIMRISK, it is possible to disaggregate the influences of urban exposure and effects from UHI and urban sensitivity, including their interaction with exposure.

In what follows we explain the procedure to compute urban and non-urban SCC for the R and RU DFs. Note that the same decomposition applies to the RP and RPU DFs. First, the total SCC value ($SCC^{RU}$) is the sum of the non-urban $SCC^{nu,R}$ and urban $SCC^{u,RU}$ values. In the $SCC^{xx,YY}$ terms, *xx* refers to the type of grid cell with *nu* and *u* denoting non-urban and urban, respectively, and *YY* refers to the type of DF used *R* (without urban effects) and *RU* (with urban effects). As such, $SCC^{nu,R}$ represents the SCC from non-urban grid cells using the R DF (i.e., no urban effects are included), and $SCC^{u,RU}$ is the SCC from urban cells when the UHI effect is considered (RU DF). This can be expressed by the following equation:

$$SCC^{RU} = SCC^{nu,R} + SCC^{u,RU} \qquad (1)$$

the urban SCC estimates ($SCC^{u,RU}$) can be further decomposed in: 1) the contribution of exposure and; 2) the combined effects of hazard (including UHI), the differences in sensitivity of the urban area to changes in climate and the interaction of these factors with exposure. The contribution of increased exposure in urban areas is obtained by calculating the SCC in urban areas without considering damages from urban effects (i.e., using the R DF) and is denoted by $SCC^{u,R}$. The contribution of the urban effects (including the UHI) plus interactions is $SCC^{u,RU^-} = [SCC^{u,RU} - SCC^{u,R}]$, the difference between the SCC value from urban grids considering the damages from urban effects $SCC^{u,RU}$ and $SCC^{u,R}$. $SCC^{u,RU^-}$ is referred to as *UHI+int* in Table 2. The complete decomposition is given by:

$$SCC^{t,RU} = SCC^{nu,R} + SCC^{u,RU} = SCC^{nu,R} + SCC^{u,R} + SCC^{u,RU^-} \qquad (2)$$

This decomposition allows us to compare the role of exposure concentration in cities under global climate change with the contribution of local climate change and urban sensitivity, within the context of global climate change (i.e., allowing for their interactions and those with exposure concentration). Note that both $SCC^{nu,R}$ and $SCC^{u,R}$ share the same damage function (sensitivity) and that, in both cases, damages are driven by global climate change alone. Differences in damages are solely due to dissimilarities in exposure and hazard levels from global warming due to location.

Finally, we define $SCC^{d,R} = SCC^{u,R} - SCC^{nu,R}$ as the change in SCC produced by an additional tonne of CO2 between urban and non-urban context derived only from differences in exposure and global climate change. $SCC^{d,R}$ is referred to as *Exposure* in Table 2. In contrast, $SCC^{u,RU^-}$ in equation (2) represents the increase in SCC due to urban characteristics including local climate change (UHI), vulnerability/sensitivity, and the interactions between local and global climate change, as well as exposure. This information can be of use for decision-makers at the city level to better understand and address local and global climate change impacts, vulnerability, and adaptation. The proposed separation can be readily applied to the RPU functions.

*The social cost of the urban heat island*

Acknowledging the relevance of the UHI effect on the economic costs of climate change, we propose a new metric called the Social Cost of the Urban Heat Island (SCUHI). This metric evaluates the variation in the present value over this century of damages from local and global climate change produced by decreasing the UHI by 1%, in all urban areas for the remaining of this century.

Analytically, the SCUHI is obtained as follows. Taking the partial derivative of the DF for urban areas described above $D_t = \alpha_R T_{GHG,t} + 2\alpha_U T_{GHG,t} T_{UHI,t} + \alpha_U T_{UHI,t}^2$ results in:

$$\frac{\partial D_t}{\partial T_{UHI}} = 2\alpha_U (T_{GHG,t} + T_{UHI,t})$$

The present value of the damages associated with a marginal change in $T_{UHI}$ is:

$$\sum_{t=0}^{\infty} \delta^t \cdot \frac{\partial D_t}{\partial T_{UHI}}$$

Moreover, since $T_{UHI}$ is modelled as $T_{UHI} = aP^b$, where $a$ and $b$ are fixed parameters and $P$ is the urban population count[44,69], SCUHI can be defined both in terms of changes in mitigation of the UHI and urban population. We first consider the impact, at the margin, of measures to reduce the urban heat island effect. Let's assume such measures affect $a$. Then,

$$\frac{\partial D_t}{\partial a} = \frac{\partial D_t}{\partial T_{UHI,t}} \cdot \frac{\partial T_{UHI,t}}{\partial a} = 2\alpha_U (T_{GHG,t} + T_{UHI,t}) P^b$$

And the SCUHI is:

$$SCUHI = \sum_{t=0}^{\infty} \delta^t \cdot \frac{\partial D_t}{\partial a}$$

This is how the SCUHI is calculated for the estimates presented in this paper, as what we propose relates to the benefits of mitigating the UHI. The SCUHI values are reported per urban dweller.

Alternatively, the SCUHI can be defined in terms of marginal changes in urban population:

$$\frac{\partial D_t}{\partial P_0} = \frac{\partial D_t}{\partial T_{UHI,t}} \cdot \frac{\partial T_{UHI,t}}{\partial P_0} = 2\alpha_U ab (T_{GHG,t} + T_{UHI,t}) P^{b-1}$$

Note that we assumed that the effects of urban population growth have a permanent effect on the urban built-up and, thus, in the local energy imbalance and in the local warming. SCUHI is then the present value of the damages associated with a marginal change in $P$:

$$SCUHI = \sum_{t=0}^{\infty} \delta^t \cdot \frac{\partial D_t}{\partial P_0}$$

Note that the SCUHI is different for each urban conurbation. The $P$ in the above equations is the population of particular cities, rather than the global population. Similarly, greenhouse warming is different in different locations.

**Author contributions:** FE, VL, WB and RSJT contributed equally to this work.

**Data availability statement**: Socioeconomic datasets and CLIMRISK output are available at https://datapincc.unam.mx/datapincc/#.

**Competing Interests Statement:** The authors declare no competing interests.

**Code availability statement:** The code for reproducing results from CLIMRISK output is available upon request.

**References**

1. IPCC. *Mitigation Pathways Compatible with 1.5°C in the Context of Sustainable Development*.

*Supplementary Information*

*Supplementary Figures*

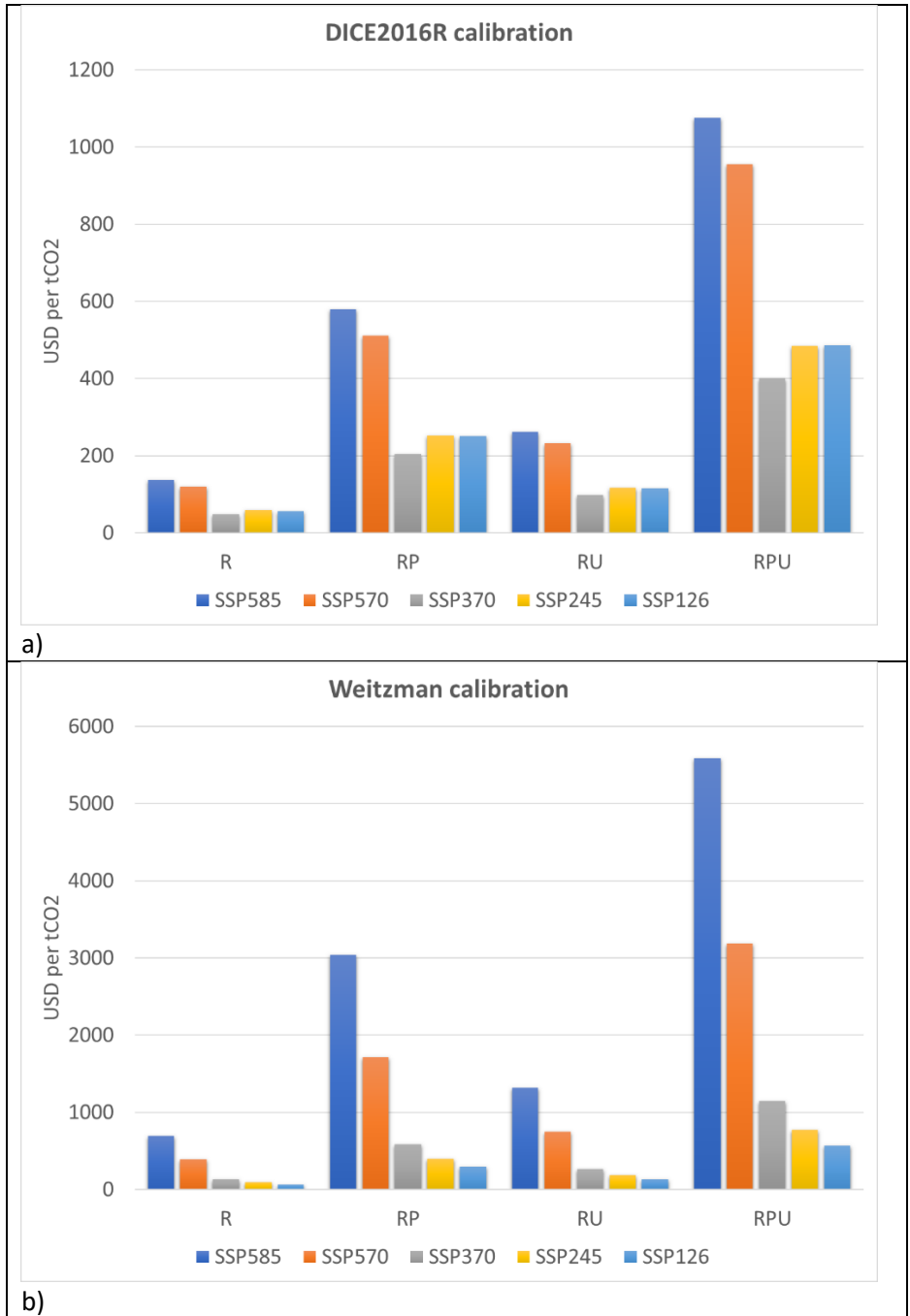

Figure S1. Global SCC values for different SSP scenarios and for calibrations based on two global damage functions. Panel a) shows the SCC values for five SSP scenarios (SSP585-blue, SSP570-orange, SSP370-grey, SSP245-yellow, and SSP126-light blue) using the RICE2016R global damage function for base calibration. Panel b) is as in a) but using the Weitzman global damage function for base calibration.

*Supplementary Tables*

**Table S1: Social cost of carbon for the SSP570 and different damage functions in CLIMRISK.**

| Region | R | RP | RU | RPU |
|---|---|---|---|---|
| US | $10.57 (8.81%) | $24.99 (4.89%) | $22.47 (9.67%) | $53.15 (5.56%) |
| EU | $12.38 (10.32%) | $23.85 (4.67%) | $25.4 (10.93%) | $48.98 (5.13%) |
| Japan | $2.24 (1.86%) | $5.32 (1.04%) | $4.62 (1.99%) | $11.01 (1.15%) |
| Russia | $1.50 (1.25%) | $6.64 (1.3%) | $4.35 (1.87%) | $19.25 (2.01%) |
| Eurasia | $1.42 (1.18%) | $4.38 (0.86%) | $3.21 (1.38%) | $9.95 (1.04%) |
| China | $15.51 (12.93%) | $69.41 (13.59%) | $40.05 (17.23%) | $179.44 (18.78%) |
| India | $22.17 (18.48%) | $96.28 (18.84%) | $39.25 (16.89%) | $170.37 (17.83%) |
| MEAST | $5.60 (4.67%) | $24.43 (4.78%) | $11.17 (4.81%) | $48.75 (5.1%) |
| Africa | $24.40 (20.34%) | $169.32 (33.14%) | $37.11 (15.97%) | $257.92 (26.99%) |
| LAM | $5.39 (4.5%) | $14.46 (2.83%) | $10.48 (4.51%) | $28.14 (2.94%) |
| OHI | $3.59 (3.00%) | $8.50 (1.66%) | $7.52 (3.23%) | $17.79 (1.86%) |
| OASIA | $13.75 (11.46%) | $59.46 (11.64%) | $23.75 (10.22%) | $102.58 (10.73%) |
| MX | $1.46 (1.21%) | $3.91 (0.76%) | $3.06 (1.32%) | $8.21 (0.86%) |
| WORLD | $119.97 (100%) | $510.93 (100%) | $232.44 (100%) | $955.54 (100%) |

Numbers in parenthesis indicate the fraction of global SCC for each region. Figures are expressed in US2005 dollars. 1.5% discount rate. DICE2016 global DF.

**Table S2: Social cost of carbon for the SSP370 and different damage functions in CLIMRISK.**

| Region | R | RP | RU | RPU |
|---|---|---|---|---|
| US | $4.43 (9.10%) | $10.67 (5.23%) | $8.87 (9.04%) | $21.4 (5.34%) |
| EU | $5.24 (10.76%) | $10.21 (5.01%) | $10.86 (11.07%) | $21.19 (5.28%) |
| Japan | $0.94 (1.93%) | $2.27 (1.11%) | $2.03 (2.07%) | $4.92 (1.23%) |
| Russia | $0.82 (1.68%) | $3.65 (1.79%) | $2.47 (2.52%) | $11.03 (2.75%) |
| Eurasia | $0.80 (1.65%) | $2.50 (1.22%) | $1.93 (1.97%) | $6.01 (1.50%) |
| China | $7.36 (15.11%) | $33.26 (16.31%) | $19.46 (19.82%) | $88.04 (21.95%) |
| India | $7.79 (16.00%) | $34.52 (16.93%) | $14.00 (14.26%) | $61.98 (15.45%) |
| MEAST | $3.06 (6.28%) | $13.56 (6.65%) | $6.33 (6.44%) | $28.03 (6.99%) |
| Africa | $7.68 (15.77%) | $56.00 (27.47%) | $12.32 (12.54%) | $89.71 (22.37%) |
| LAM | $2.83 (5.81%) | $7.63 (3.74%) | $5.69 (5.80%) | $15.39 (3.84%) |
| OHI | $1.46 (3.00%) | $3.51 (1.72%) | $2.88 (2.93%) | $6.93 (1.73%) |
| OASIA | $5.39 (11.08%) | $23.72 (11.63%) | $9.39 (9.57%) | $41.23 (10.28%) |
| MX | $0.89 (1.83%) | $2.41 (1.18%) | $1.94 (1.97%) | $5.23 (1.30%) |
| WORLD | $48.68 (100%) | $203.91 (100%) | $98.17 (100%) | $401.08 (100%) |

Numbers in parenthesis indicate the fraction of global SCC for each region. Figures are expressed in US2005 dollars. 1.5% discount rate. DICE2016 global DF.

**Table S3: Social cost of carbon for the SSP245 and different damage functions in CLIMRISK.**

| Region | R | RP | RU | RPU |
|---|---|---|---|---|
| US | $4.82 (8.20%) | $11.58 (4.60%) | $10.41 (8.87%) | $25.03 (5.17%) |
| EU | $6.19 (10.53%) | $12.03 (4.78%) | $13.28 (11.31%) | $25.82 (5.33%) |
| Japan | $1.08 (1.83%) | $2.59 (1.03%) | $2.38 (2.02%) | $5.73 (1.18%) |
| Russia | $0.81 (1.38%) | $3.64 (1.44%) | $2.42 (2.06%) | $10.86 (2.24%) |
| Eurasia | $0.79 (1.34%) | $2.46 (0.98%) | $1.88 (1.60%) | $5.86 (1.21%) |
| China | $8.28 (14.09%) | $37.44 (14.87%) | $22.05 (18.79%) | $99.70 (20.59%) |
| India | $10.95 (18.62%) | $48.14 (19.12%) | $19.47 (16.58%) | $85.53 (17.66%) |
| MEAST | $3.12 (5.31%) | $13.84 (5.50%) | $6.29 (5.36%) | $27.85 (5.75%) |
| Africa | $10.79 (18.35%) | $77.01 (30.59%) | $16.69 (14.22%) | $119.01 (24.58%) |
| LAM | $2.98 (5.06%) | $8.04 (3.19%) | $6.09 (5.18%) | $16.45 (3.40%) |
| OHI | $1.66 (2.83%) | $3.99 (1.59%) | $3.59 (3.06%) | $8.62 (1.78%) |
| OASIA | $6.67 (11.34%) | $29.21 (11.6%) | $11.39 (9.70%) | $49.84 (10.29%) |
| MX | $0.66 (1.12%) | $1.78 (0.71%) | $1.46 (1.25%) | $3.96 (0.82%) |
| WORLD | $58.81 (100%) | $251.75 (100%) | $117.39 (100%) | $484.27 (100%) |

Numbers in parenthesis indicate the fraction of global SCC for each region. Figures are expressed in US2005 dollars. 1.5% discount rate. DICE2016 global DF.

**Table S4: Social cost of carbon for the SSP126 and different damage functions in CLIMRISK.**

| Region | R | RP | RU | RPU |
|---|---|---|---|---|
| US | $4.43 (7.81%) | $10.66 (4.25%) | $9.77 (8.53%) | $23.52 (4.84%) |
| EU | $5.27 (9.29%) | $10.27 (4.09%) | $11.68 (10.19%) | $22.77 (4.69%) |
| Japan | $1.05 (1.84%) | $2.52 (1.01%) | $2.41 (2.10%) | $5.81 (1.20%) |
| Russia | $0.69 (1.22%) | $3.14 (1.25%) | $2.13 (1.86%) | $9.70 (2.00%) |
| Eurasia | $0.69 (1.23%) | $2.18 (0.87%) | $1.67 (1.46%) | $5.26 (1.08%) |
| China | $8.40 (14.81%) | $38.30 (15.27%) | $22.66 (19.77%) | $103.29 (21.25%) |
| India | $11.29 (19.92%) | $50.29 (20.05%) | $19.88 (17.35%) | $88.52 (18.22%) |
| MEAST | $2.64 (4.65%) | $11.84 (4.72%) | $5.32 (4.64%) | $23.85 (4.91%) |
| Africa | $10.82 (19.09%) | $79.76 (31.8%) | $16.73 (14.60%) | $123.35 (25.38%) |
| LAM | $2.69 (4.75%) | $7.32 (2.92%) | $5.71 (4.99%) | $15.54 (3.20%) |
| OHI | $1.43 (2.52%) | $3.44 (1.37%) | $3.16 (2.76%) | $7.61 (1.57%) |
| OASIA | $6.53 (11.51%) | $29.03 (11.57%) | $11.69 (10.20%) | $51.87 (10.67%) |
| MX | $0.77 (1.36%) | $2.09 (0.83%) | $1.78 (1.56%) | $4.85 (1.00%) |
| WORLD | $56.71 (100%) | $250.85 (100%) | $114.59 (100%) | $485.94 (100%) |

Numbers in parenthesis indicate the fraction of global SCC for each region. Figures are expressed in US2005 dollars. 1.5% discount rate. DICE2016 global DF.

**Table S5: Social cost of carbon for the SSP585 and different damage functions in CLIMRISK.**

| Region | R | RP | RU | RPU |
|---|---|---|---|---|
| US | $57.96 (8.4%) | $136.71 (4.5%) | $125.24 (9.49%) | $295.62 (5.29%) |
| EU | $68.44 (9.92%) | $131.65 (4.33%) | $142.28 (10.79%) | $274.05 (4.9%) |
| Japan | $11.25 (1.63%) | $26.74 (0.88%) | $23.18 (1.76%) | $55.23 (0.99%) |
| Russia | $7.33 (1.06%) | $32.66 (1.07%) | $21.19 (1.61%) | $94.91 (1.7%) |
| Eurasia | $7.17 (1.04%) | $22.28 (0.73%) | $16.25 (1.23%) | $50.64 (0.91%) |
| China | $69.44 (10.06%) | $317.34 (10.44%) | $178.85 (13.56%) | $820.3 (14.67%) |
| India | $131.39 (19.04%) | $576.1 (18.95%) | $234.09 (17.75%) | $1026.29 (18.36%) |
| MEAST | $30.94 (4.48%) | $135.51 (4.46%) | $62.03 (4.7%) | $272.2 (4.87%) |
| Africa | $168.39 (24.4%) | $1162.25 (38.23%) | $256.98 (19.48%) | $1779.1 (31.83%) |
| LAM | $28.97 (4.2%) | $77.94 (2.56%) | $56.62 (4.29%) | $152.75 (2.73%) |
| OHI | $19.95 (2.89%) | $47.1 (1.55%) | $42.27 (3.2%) | $99.95 (1.79%) |
| OASIA | $81.14 (11.76%) | $352.7 (11.6%) | $143.63 (10.89%) | $624.32 (11.17%) |
| MX | $7.82 (1.13%) | $21 (0.69%) | $16.51 (1.25%) | $44.48 (0.8%) |
| WORLD | $690.2 (100%) | $3039.99 (100%) | $1319.14 (100%) | $5589.85 (100%) |

Numbers in parenthesis indicate the fraction of global SCC for each region. Figures are expressed in US2005 dollars. 1.5% discount rate. Weitzman global DF.

**Table S5: Social cost of carbon for the SSP570 and different damage functions in CLIMRISK.**

| Region | R | RP | RU | RPU |
|---|---|---|---|---|
| US | $31.86 (8.25%) | $75.34 (4.4%) | $70.56 (9.42%) | $166.99 (5.24%) |
| EU | $37.55 (9.73%) | $72.36 (4.23%) | $80.55 (10.75%) | $155.4 (4.88%) |
| Japan | $6.29 (1.63%) | $14.98 (0.87%) | $13.43 (1.79%) | $32.06 (1.01%) |
| Russia | $4.12 (1.07%) | $18.44 (1.08%) | $12.21 (1.63%) | $54.92 (1.72%) |
| Eurasia | $4 (1.04%) | $12.46 (0.73%) | $9.24 (1.23%) | $28.84 (0.91%) |
| China | $40.44 (10.47%) | $185.11 (10.81%) | $105.78 (14.12%) | $485.78 (15.25%) |
| India | $74.97 (19.41%) | $330.09 (19.28%) | $133.41 (17.8%) | $587.38 (18.44%) |
| MEAST | $17.66 (4.57%) | $77.74 (4.54%) | $35.59 (4.75%) | $156.94 (4.93%) |
| Africa | $92.68 (24%) | $647.24 (37.81%) | $142.03 (18.96%) | $994.97 (31.23%) |
| LAM | $16.17 (4.19%) | $43.59 (2.55%) | $32.49 (4.34%) | $87.83 (2.76%) |
| OHI | $10.94 (2.83%) | $25.88 (1.51%) | $23.81 (3.18%) | $56.42 (1.77%) |
| OASIA | $45.11 (11.68%) | $197.03 (11.51%) | $80.71 (10.77%) | $352.54 (11.07%) |
| MX | $4.37 (1.13%) | $11.76 (0.69%) | $9.5 (1.27%) | $25.66 (0.81%) |
| WORLD | $386.16 (100%) | $1712.02 (100%) | $749.3 (100%) | $3185.73 (100%) |

Numbers in parenthesis indicate the fraction of global SCC for each region. Figures are expressed in US2005 dollars. 1.5% discount rate. Weitzman global DF.

**Table S6: Social cost of carbon for the SSP370 and different damage functions in CLIMRISK.**

| Region | R | RP | RU | RPU |
|---|---|---|---|---|
| US | $10.11 (7.67%) | $24.54 (4.22%) | $20.2 (7.56%) | $49.12 (4.28%) |
| EU | $12.26 (9.31%) | $23.99 (4.13%) | $25.88 (9.69%) | $50.7 (4.42%) |
| Japan | $1.92 (1.46%) | $4.68 (0.81%) | $4.22 (1.58%) | $10.3 (0.9%) |
| Russia | $2.07 (1.57%) | $9.33 (1.61%) | $6.49 (2.43%) | $29.25 (2.55%) |
| Eurasia | $2.12 (1.61%) | $6.62 (1.14%) | $5.27 (1.97%) | $16.43 (1.43%) |
| China | $17.31 (13.14%) | $79.81 (13.74%) | $46.49 (17.4%) | $214.85 (18.74%) |
| India | $22.6 (17.15%) | $101.36 (17.45%) | $41.4 (15.5%) | $185.49 (16.18%) |
| MEAST | $8.57 (6.5%) | $38.35 (6.6%) | $18.24 (6.83%) | $81.56 (7.12%) |
| Africa | $25.65 (19.46%) | $187.17 (32.23%) | $42.2 (15.8%) | $307.48 (26.82%) |
| LAM | $7.6 (5.77%) | $20.6 (3.55%) | $15.78 (5.91%) | $42.87 (3.74%) |
| OHI | $3.4 (2.58%) | $8.23 (1.42%) | $6.72 (2.52%) | $16.31 (1.42%) |
| OASIA | $15.72 (11.93%) | $69.53 (11.97%) | $28.76 (10.77%) | $127.14 (11.09%) |
| MX | $2.43 (1.84%) | $6.57 (1.13%) | $5.47 (2.05%) | $14.85 (1.3%) |
| WORLD | $131.78 (100%) | $580.8 (100%) | $267.12 (100%) | $1146.35 (100%) |

Numbers in parenthesis indicate the fraction of global SCC for each region. Figures are expressed in US2005 dollars. 1.5% discount rate. Weitzman global DF.

**Table S7: Social cost of carbon for the SSP245 and different damage functions in CLIMRISK.**

| Region | R | RP | RU | RPU |
|---|---|---|---|---|
| US | $6.84 (7.54%) | $16.5 (4.13%) | $15.14 (8.28%) | $36.56 (4.73%) |
| EU | $8.95 (9.86%) | $17.44 (4.37%) | $19.82 (10.84%) | $38.63 (5%) |
| Japan | $1.49 (1.64%) | $3.59 (0.9%) | $3.38 (1.85%) | $8.18 (1.06%) |
| Russia | $1.17 (1.29%) | $5.29 (1.32%) | $3.57 (1.95%) | $16.19 (2.09%) |
| Eurasia | $1.16 (1.27%) | $3.62 (0.91%) | $2.81 (1.54%) | $8.81 (1.14%) |
| China | $11.91 (13.12%) | $54.53 (13.65%) | $32.22 (17.62%) | $147.69 (19.11%) |
| India | $17.76 (19.56%) | $78.68 (19.7%) | $31.81 (17.4%) | $140.93 (18.24%) |
| MEAST | $4.84 (5.33%) | $21.6 (5.41%) | $9.88 (5.4%) | $44.12 (5.71%) |
| Africa | $18.28 (20.13%) | $131.07 (32.82%) | $28.68 (15.69%) | $205.55 (26.6%) |
| LAM | $4.47 (4.92%) | $12.12 (3.03%) | $9.44 (5.16%) | $25.64 (3.32%) |
| OHI | $2.4 (2.64%) | $5.77 (1.44%) | $5.31 (2.91%) | $12.8 (1.66%) |
| OASIA | $10.55 (11.62%) | $46.48 (11.64%) | $18.52 (10.13%) | $81.59 (10.56%) |
| MX | $0.98 (1.08%) | $2.66 (0.67%) | $2.26 (1.23%) | $6.13 (0.79%) |
| WORLD | $90.79 (100%) | $399.36 (100%) | $182.84 (100%) | $772.82 (100%) |

Numbers in parenthesis indicate the fraction of global SCC for each region. Figures are expressed in US2005 dollars. 1.5% discount rate. Weitzman global DF.

**Table S8: Social cost of carbon for the SSP126 and different damage functions in CLIMRISK.**

| Region | R | RP | RU | RPU |
|---|---|---|---|---|
| US | $5.02 (7.63%) | $12.09 (4.13%) | $11.22 (8.37%) | $27.05 (4.73%) |
| EU | $5.97 (9.08%) | $11.65 (3.98%) | $13.48 (10.06%) | $26.31 (4.6%) |
| Japan | $1.18 (1.79%) | $2.85 (0.97%) | $2.77 (2.07%) | $6.69 (1.17%) |
| Russia | $0.78 (1.19%) | $3.57 (1.22%) | $2.45 (1.83%) | $11.20 (1.96%) |
| Eurasia | $0.79 (1.20%) | $2.49 (0.85%) | $1.93 (1.44%) | $6.07 (1.06%) |
| China | $9.63 (14.64%) | $44.06 (15.04%) | $26.21 (19.57%) | $120.03 (21.01%) |
| India | $13.34 (20.28%) | $59.52 (20.31%) | $23.56 (17.59%) | $105.14 (18.41%) |
| MEAST | $3.07 (4.66%) | $13.8 (4.71%) | $6.22 (4.64%) | $27.98 (4.90%) |
| Africa | $12.78 (19.44%) | $94.33 (32.2%) | $19.86 (14.83%) | $146.66 (25.67%) |
| LAM | $3.09 (4.7%) | $8.42 (2.87%) | $6.70 (5.00%) | $18.26 (3.20%) |
| OHI | $1.62 (2.46%) | $3.90 (1.33%) | $3.64 (2.72%) | $8.78 (1.54%) |
| OASIA | $7.61 (11.57%) | $33.90 (11.57%) | $13.8 (10.3%) | $61.38 (10.74%) |
| MX | $0.88 (1.35%) | $2.41 (0.82%) | $2.09 (1.56%) | $5.70 (1.00%) |
| WORLD | $65.76 (100%) | $292.98 (100%) | $133.93 (100%) | $571.24 (100%) |

Numbers in parenthesis indicate the fraction of global SCC for each region. Figures are expressed in US2005 dollars. 1.5% discount rate. Weitzman global DF.

**Table S9: Social cost of carbon for the SSP585, different damage functions in CLIMRISK and a 3% discount rate.**

| Region | R | RP | RU | RPU |
|---|---|---|---|---|
| US | $5.38 (9.17%) | $12.60 (5.27%) | $11.37 (9.92%) | $26.66 (5.9%) |
| EU | $6.27 (10.7%) | $12.01 (5.03%) | $12.84 (11.2%) | $24.6 (5.44%) |
| Japan | $1.18 (2.02%) | $2.79 (1.17%) | $2.46 (2.15%) | $5.81 (1.28%) |
| Russia | $0.79 (1.35%) | $3.4 (1.42%) | $2.29 (2%) | $9.85 (2.18%) |
| Eurasia | $0.73 (1.25%) | $2.23 (0.93%) | $1.66 (1.45%) | $5.06 (1.12%) |
| China | $8.33 (14.21%) | $36.06 (15.09%) | $21.52 (18.79%) | $93.24 (20.62%) |
| India | $10.49 (17.89%) | $44.41 (18.58%) | $18.57 (16.21%) | $78.6 (17.38%) |
| MEAST | $2.78 (4.75%) | $11.83 (4.95%) | $5.54 (4.84%) | $23.6 (5.22%) |
| Africa | $10.8 (18.43%) | $72.45 (30.32%) | $16.43 (14.34%) | $110.38 (24.41%) |
| LAM | $2.72 (4.64%) | $7.21 (3.02%) | $5.30 (4.62%) | $14.04 (3.1%) |
| OHI | $1.81 (3.09%) | $4.24 (1.78%) | $3.77 (3.29%) | $8.84 (1.95%) |
| OASIA | $6.58 (11.23%) | $27.78 (11.62%) | $11.25 (9.82%) | $47.45 (10.49%) |
| MX | $0.74 (1.26%) | $1.96 (0.82%) | $1.55 (1.36%) | $4.12 (0.91%) |
| WORLD | $58.61 (100%) | $238.98 (100%) | $114.57 (100%) | $452.26 (100%) |

Numbers in parenthesis indicate the fraction of global SCC for each region. Figures are expressed in US2005 dollars. 3% discount rate. DICE2016 global DF.

**Table S10: Social cost of carbon for the SSP370, different damage functions in CLIMRISK and a 3% discount rate.**

| Region | R | RP | RU | RPU |
|---|---|---|---|---|
| US | $2.36 (9.79%) | $5.62 (5.83%) | $4.88 (9.84%) | $11.62 (6.02%) |
| EU | $2.77 (11.49%) | $5.36 (5.56%) | $5.9 (11.89%) | $11.41 (5.91%) |
| Japan | $0.54 (2.22%) | $1.28 (1.33%) | $1.20 (2.42%) | $2.86 (1.48%) |
| Russia | $0.41 (1.71%) | $1.79 (1.86%) | $1.26 (2.54%) | $5.47 (2.83%) |
| Eurasia | $0.39 (1.63%) | $1.20 (1.25%) | $0.95 (1.92%) | $2.91 (1.51%) |
| China | $3.88 (16.08%) | $16.99 (17.61%) | $10.38 (20.93%) | $45.41 (23.52%) |
| India | $3.78 (15.64%) | $16.31 (16.91%) | $6.75 (13.62%) | $29.14 (15.09%) |
| MEAST | $1.50 (6.22%) | $6.50 (6.73%) | $3.10 (6.24%) | $13.38 (6.93%) |
| Africa | $3.37 (13.94%) | $23.79 (24.66%) | $5.39 (10.87%) | $38.05 (19.7%) |
| LAM | $1.39 (5.75%) | $3.70 (3.84%) | $2.86 (5.76%) | $7.63 (3.95%) |
| OHI | $0.77 (3.18%) | $1.82 (1.89%) | $1.55 (3.13%) | $3.70 (1.91%) |
| OASIA | $2.54 (10.53%) | $10.93 (11.33%) | $4.41 (8.9%) | $18.94 (9.81%) |
| MX | $0.44 (1.80%) | $1.16 (1.20%) | $0.96 (1.95%) | $2.57 (1.33%) |
| WORLD | $24.14 (100%) | $96.46 (100%) | $49.6 (100%) | $193.09 (100%) |

Numbers in parenthesis indicate the fraction of global SCC for each region. Figures are expressed in US2005 dollars. 3% discount rate. DICE2016 global DF.

**Table S11: Social cost of carbon for the SSP245, different damage functions in CLIMRISK and a 3% discount rate.**

| Region | R | RP | RU | RPU |
|---|---|---|---|---|
| US | $2.53 (8.97%) | $6.01 (5.23%) | $5.56 (9.67%) | $13.21 (5.85%) |
| EU | $3.17 (11.21%) | $6.11 (5.31%) | $6.9 (11.99%) | $13.32 (5.90%) |
| Japan | $0.59 (2.1%) | $1.41 (1.23%) | $1.35 (2.34%) | $3.21 (1.42%) |
| Russia | $0.41 (1.47%) | $1.80 (1.57%) | $1.25 (2.18%) | $5.46 (2.42%) |
| Eurasia | $0.39 (1.39%) | $1.20 (1.05%) | $0.94 (1.63%) | $2.88 (1.27%) |
| China | $4.33 (15.34%) | $18.92 (16.46%) | $11.63 (20.21%) | $50.77 (22.49%) |
| India | $5.05 (17.89%) | $21.67 (18.85%) | $8.96 (15.58%) | $38.4 (17.01%) |
| MEAST | $1.53 (5.41%) | $6.60 (5.74%) | $3.07 (5.34%) | $13.26 (5.87%) |
| Africa | $4.52 (16.02%) | $31.35 (27.27%) | $6.99 (12.15%) | $48.38 (21.44%) |
| LAM | $1.46 (5.16%) | $3.88 (3.38%) | $3.04 (5.28%) | $8.10 (3.59%) |
| OHI | $0.85 (3.02%) | $2.02 (1.76%) | $1.87 (3.26%) | $4.44 (1.97%) |
| OASIA | $3.06 (10.84%) | $13.10 (11.39%) | $5.21 (9.07%) | $22.3 (9.88%) |
| MX | $0.33 (1.16%) | $0.88 (0.76%) | $0.75 (1.30%) | $1.99 (0.88%) |
| WORLD | $28.24 (100%) | $114.96 (100%) | $57.52 (100%) | $225.71 (100%) |

Numbers in parenthesis indicate the fraction of global SCC for each region. Figures are expressed in US2005 dollars. 3% discount rate. DICE2016 global DF.

**Table S12: Social cost of carbon for the SSP126, different damage functions in CLIMRISK and a 3% discount rate.**

| Region | R | RP | RU | RPU |
|---|---|---|---|---|
| US | $2.41 (8.54%) | $5.72 (4.86%) | $5.36 (9.24%) | $12.74 (5.48%) |
| EU | $2.84 (10.07%) | $5.49 (4.67%) | $6.34 (10.92%) | $12.25 (5.27%) |
| Japan | $0.59 (2.1%) | $1.41 (1.2%) | $1.38 (2.38%) | $3.29 (1.41%) |
| Russia | $0.38 (1.34%) | $1.65 (1.41%) | $1.17 (2.02%) | $5.13 (2.21%) |
| Eurasia | $0.37 (1.3%) | $1.12 (0.96%) | $0.88 (1.52%) | $2.71 (1.17%) |
| China | $4.56 (16.16%) | $19.98 (16.99%) | $12.35 (21.28%) | $54.12 (23.28%) |
| India | $5.37 (19.03%) | $23.18 (19.71%) | $9.44 (16.26%) | $40.75 (17.53%) |
| MEAST | $1.36 (4.81%) | $5.89 (5.01%) | $2.73 (4.71%) | $11.87 (5.11%) |
| Africa | $4.69 (16.62%) | $33.16 (28.19%) | $7.25 (12.5%) | $51.33 (22.08%) |
| LAM | $1.38 (4.89%) | $3.69 (3.14%) | $2.96 (5.1%) | $7.92 (3.41%) |
| OHI | $0.77 (2.72%) | $1.82 (1.55%) | $1.71 (2.94%) | $4.06 (1.75%) |
| OASIA | $3.11 (11.04%) | $13.44 (11.43%) | $5.53 (9.53%) | $23.84 (10.25%) |
| MX | $0.4 (1.4%) | $1.06 (0.9%) | $0.93 (1.59%) | $2.48 (1.07%) |
| WORLD | $28.21 (100%) | $117.61 (100%) | $58.03 (100%) | $232.48 (100%) |

Numbers in parenthesis indicate the fraction of global SCC for each region. Figures are expressed in US2005 dollars. 3% discount rate. DICE2016 global DF.

Table S13: Contributions of non-urban areas, urban exposure, and UHI plus interactions in urban areas to regional and world SCC, under different SSP scenarios.

| US | Non-urban | Urban-exposure | UHI+int | China | Non-urban | Urban-exposure | UHI+int | OHI | Non-urban | Urban-exposure | UHI+int |
|---|---|---|---|---|---|---|---|---|---|---|---|
| SSP585 | 12.31% | 35.61% | 52.08% | SSP585 | 2.09% | 37.07% | 60.85% | SSP585 | 10.72% | 38.00% | 51.28% |
| SSP370 | 18.35% | 31.56% | 50.09% | SSP370 | 1.46% | 36.35% | 62.20% | SSP370 | 16.86% | 33.77% | 49.37% |
| SSP245 | 14.04% | 32.25% | 53.71% | SSP245 | 1.68% | 35.89% | 62.44% | SSP245 | 9.88% | 36.37% | 53.74% |
| SSP126 | 14.29% | 31.07% | 54.64% | SSP126 | 1.92% | 35.16% | 62.92% | SSP126 | 11.22% | 33.88% | 54.90% |
| EU | Non-urban | Urban-exposure | UHI+int | India | Non-urban | Urban-exposure | UHI+int | OASIA | Non-urban | Urban-exposure | UHI+int |
| SSP585 | 7.70% | 42.15% | 50.15% | SSP585 | 0.28% | 56.08% | 43.64% | SSP585 | 4.35% | 53.99% | 41.65% |
| SSP370 | 7.83% | 40.40% | 51.77% | SSP370 | 0.19% | 55.43% | 44.38% | SSP370 | 3.31% | 54.08% | 42.61% |
| SSP245 | 7.47% | 39.16% | 53.37% | SSP245 | 0.25% | 56.01% | 43.75% | SSP245 | 4.29% | 54.24% | 41.48% |
| SSP126 | 7.66% | 37.47% | 54.87% | SSP126 | 0.28% | 56.52% | 43.20% | SSP126 | 4.16% | 51.66% | 44.18% |
| Japan | Non-urban | Urban-exposure | UHI+int | MEAST | Non-urban | Urban-exposure | UHI+int | MX | Non-urban | Urban-exposure | UHI+int |
| SSP585 | 2.92% | 46.68% | 50.40% | SSP585 | 7.42% | 42.85% | 49.73% | SSP585 | 8.81% | 39.86% | 51.33% |
| SSP370 | 2.74% | 43.44% | 53.83% | SSP370 | 5.45% | 42.87% | 51.68% | SSP370 | 6.79% | 39.32% | 53.89% |
| SSP245 | 2.88% | 42.41% | 54.71% | SSP245 | 7.08% | 42.59% | 50.33% | SSP245 | 7.68% | 37.38% | 54.94% |
| SSP126 | 2.69% | 40.77% | 56.54% | SSP126 | 7.62% | 42.01% | 50.37% | SSP126 | 7.67% | 35.50% | 56.83% |
| Russia | Non-urban | Urban-exposure | UHI+int | Africa | Non-urban | Urban-exposure | UHI+int | WORLD | Non-urban | Urban-exposure | UHI+int |
| SSP585 | 11.07% | 24.12% | 64.81% | SSP585 | 11.67% | 53.84% | 34.48% | SSP585 | 6.60% | 45.49% | 47.91% |
| SSP370 | 9.40% | 23.72% | 66.89% | SSP370 | 6.60% | 55.42% | 37.98% | SSP370 | 6.04% | 43.50% | 50.46% |
| SSP245 | 10.86% | 22.67% | 66.48% | SSP245 | 9.22% | 55.05% | 35.72% | SSP245 | 6.00% | 44.04% | 49.95% |
| SSP126 | 10.61% | 21.77% | 67.63% | SSP126 | 11.24% | 53.10% | 35.66% | SSP126 | 6.26% | 43.18% | 50.56% |
| Eurasia | Non-urban | Urban-exposure | UHI+int | LAM | Non-urban | Urban-exposure | UHI+int | | | | |
| SSP585 | 15.27% | 29.44% | 55.28% | SSP585 | 11.66% | 40.82% | 47.52% | | | | |
| SSP370 | 11.89% | 29.67% | 58.44% | SSP370 | 9.46% | 40.17% | 50.38% | | | | |
| SSP245 | 13.46% | 28.63% | 57.91% | SSP245 | 10.65% | 38.22% | 51.13% | | | | |
| SSP126 | 14.20% | 27.34% | 58.46% | SSP126 | 10.42% | 36.67% | 52.90% | | | | |

Numbers are in percent of total SCC and add up to 100%.

**Table S15: Relation between world regions and countries included in CLIMRISK.**

| Region | Countries |
|---|---|
| USA | USA, Puerto Rico |
| EU | Austria, Belgium, Channel Islands, Cyprus, Czech Republic, Denmark, Estonia, Faeroe Islands, Finland, France, Germany, Greece, Greenland, Hungary, Iceland, Ireland, Italy, Luxembourg, Malta, Netherlands, Norway, Poland, Portugal, Slovakia, Slovenia, Spain, Sweden, Switzerland, Turkey, United Kingdom |
| JAPAN | Japan |
| RUSSIA | Russia |
| EURASIA | Albania, Bosnia and Herzegovina, Bulgaria, Croatia, Romania, TFYR Macedonia, Yugoslavia, Azerbaijan, Armenia, Belarus, Georgia, Kazakhstan, Kyrgyzstan, Latvia, Lithuania, Republic of Moldova, Tajikistan, Turkmenistan, Ukraine, Uzbekistan |
| CHINA | China, Hong Kong SAR |
| INDIA | India |
| MEAST | Bahrain, Occupied Palestinian Terr., Iran (Islamic Republic of), Iraq, Israel, Jordan, Kuwait, Lebanon, Oman, Qatar, Saudi Arabia, Syrian Arab Republic, United Arab Emirates, Yemen |
| AFRICA | Angola, Botswana, Burundi, Cameroon, Cape Verde, Central African Republic, Chad, Comoros, Congo, Dem. Rep. of the Congo, Benin, Equatorial Guinea, Ethiopia, Eritrea, Djibouti, Gabon, Gambia, Ghana, Guinea, Cote d'Ivoire, Kenya, Lesotho, Liberia, Madagascar, Malawi, Mali, Mauritania, Mozambique, Namibia, Niger, Nigeria, Guinea-Bissau, Reunion, Rwanda, Senegal, Sierra Leone, Somalia, South Africa, Zimbabwe, Swaziland, Togo, Uganda, United Republic of Tanzania, Burkina Faso, Zambia, Algeria, Libyan Arab Jamahiriya, Morocco, Western Sahara, Sudan, Tunisia, Egypt |
| LAM | Argentina, Bahamas, Barbados, Bolivia, Brazil, Belize, Chile, Colombia, Costa Rica, Cuba, Dominican Republic, Ecuador, El Salvador, French Guiana, Grenada, Guadeloupe, Guatemala, Guyana, Haiti, Honduras, Jamaica, Martinique, Netherlands Antilles, Nicaragua, Panama, Paraguay, Peru, Saint Vincent and the |

|  | Grenadines, Suriname, Trinidad and Tobago, Uruguay, Venezuela |
|---|---|
| OHI | Australia, Canada, New Zealand, Republic of Korea, Taiwan |
| OTHASIA | Afghanistan, Bangladesh, Bhutan, Brunei Darussalam, Cambodia, Dem. People's Rep. of Korea, East Timor, Fiji, French Polynesia, Indonesia, Lao People's Dem. Republic, Malaysia, Mongolia, Myanmar, New Caledonia, Papua New Guinea, Philippines, Samoa, Singapore, Solomon Islands, Thailand, Vanuatu, Viet Nam |
| MX | Mexico |

**Table S16: List of Atmosphere-Ocean General Circulation Model (AOGCM) of the Coupled Model Intercomparison Project (CMIP6) included in CLIMRISK.**

| | | | |
|---|---|---|---|
| ACCESS-CM2 | CNRM-ESM2-1-f2 | GISS-E2-1-G-p1 | MPI-ESM1-2-HR |
| ACCESS-ESM1-5 | CanESM5-CanOE | GISS-E2-1-G-p3 | MPI-ESM1-2-LR |
| AWI-CM-1-1-MR | CanESM5-p1 | HadGEM3-GC31-LL-f3 | MRI-ESM2-0 |
| BCC-CSM2-MR | CanESM5-p2 | HadGEM3-GC31-MM-f3 | NESM3 |
| CESM2-WACCM | EC-Earth3-Veg | INM-CM4-8 | NorESM2-LM |
| CESM2 | EC-Earth3 | INM-CM5-0 | NorESM2-MM |
| CIESM | FGOALS-f3-L | IPSL-CM6A-LR | UKESM1-0-LL-f2 |
| CMCC-CM2-SR5 | FGOALS-g3 | KACE-1-0-G | |
| CNRM-CM6-1-HR-f2 | FIO-ESM-2-0 | MIROC-ES2L-f2 | |
| CNRM-CM6-1-f2 | GFDL-ESM4 | MIROC6 | |